# On the fractal structure of two-dimensional quantum gravity

J. Ambjørn, J. Jurkiewicz[1] and Y. Watabiki[2]

The Niels Bohr Institute
Blegdamsvej 17, DK-2100 Copenhagen Ø, Denmark

## Abstract

We provide evidence that the Hausdorff dimension is 4 and the spectral dimension is 2 for two-dimensional quantum gravity coupled the matter with a central charge $c \leq 1$. For $c > 1$ the Hausdorff dimension and the spectral dimension monotonously decreases to 2 and 1, respectively.

---

[1] Permanent address: Inst. of Phys., Jagellonian University., ul. Reymonta 4, PL-30 059, Kraków 16, Poland.

[2] Address after October 1st, 1995, Department of Physics, Tokyo Institute of Technology, Oh-Okayama, Meguro, Tokyo 152, Japan.

# 1 Introduction

The fractal structure of space-time is of primary interest in any theory of quantum gravity. Although the starting point usually is manifolds of a specific dimension $d$, the quantum theory instructs us to perform the average over all equivalence classes of metrics and in this way the "effective" dimension of space-time can be different from $d$. Presently we do not have a well-defined theory of quantum gravity in four dimensions. Two-dimensional gravity is an interesting laboratory, where we have available both analytical and numerical tools with which we can study the fractal structure, and in the following we will confine ourselves to two-dimensional quantum gravity, although some of the definitions and relations are given for arbitrary $d$.

We assume that the partition function for Euclidean quantum gravity can be written as:

$$Z(\Lambda) = \iint \mathcal{D}[g_{ab}]\mathcal{D}\phi \, e^{-S[g,\phi;G] - \Lambda \int \sqrt{g}}, \tag{1}$$

where the integration is over equivalence classes of metrics $[g_{ab}]$, and the action is the gravitational action, usually taken to be the Einstein-Hilbert action which depends on the gravitational constant $G$. $\phi$ symbolizes the matter fields, and the dependence of the matter coupling is suppressed. It is sometimes convenient to consider the partition function for a fixed volume $V$. Since the volume is conjugate to the cosmological constant, we can define the finite volume partition function without reference to $\Lambda$:

$$Z(V) = \iint \mathcal{D}[g_{ab}]\mathcal{D}\phi \, \delta(\int \sqrt{g} - V) \, e^{-S[g,\phi;G]}. \tag{2}$$

The partition functions in (1) and (2) are related as

$$Z(\Lambda) = \int_0^\infty dV e^{-\Lambda V} Z(V). \tag{3}$$

Recently it was shown [1, 2, 3] that the volume-volume correlator of Euclidean quantum gravity is a perfect probe of the fractal structure of space-time. It is defined as

$$G_\Lambda(R) = \iint \mathcal{D}[g_{ab}]\mathcal{D}\phi \, e^{-S[g,\phi;G] - \Lambda \int \sqrt{g}} \iint d^d\xi \sqrt{g(\xi)} \, d^d\xi' \sqrt{g(\xi')} \, \delta(d_g(\xi,\xi') - R), \tag{4}$$

where $d_g(\xi,\xi')$ denotes the geodesic distance between the points labeled by $\xi$ and $\xi'$, calculated with the metric $g_{ab}$. One can view $G_\Lambda(R)$ as the partition function for universes with two marked points separated by a geodesic distance $R$. If scaling arguments can be applied to the system, one expects a generic behavior

$$G_\Lambda(R) \sim \Lambda^{\nu - \gamma_s} \hat{F}(\Lambda^\nu R), \tag{5}$$

where $\gamma_s$ is the string susceptibility of the system, i.e.,

$$Z(\Lambda) \sim \text{const.} \, \Lambda^{2-\gamma_s} + \text{less singular terms.} \tag{6}$$

The behavior (5) follows from the definitions since

$$\frac{\partial^2 Z(\Lambda)}{\partial^2 \Lambda} = \int_0^\infty dR \, G_\Lambda(R) \sim \text{const.} \, \Lambda^{-\gamma_s} + \text{less singular terms.} \tag{7}$$

However, there is a slight subtlety associated with (5) and (7). If $-1 < \gamma_s < 0$, as is the case for unitary theories with $0 \leq c < 1$, the coefficient in front of (7) is *negative* (cf. [1]).

Since the function $\hat{F}$ in (5) is positive for unitary theories and falls of fast for large $R$ (see (18) below) this is only possible if the function $\hat{F}(x)$ is singular for small $x$, such that the term $\Lambda^{-\gamma_s}$ is a subdominant term, while the dominant term is analytic in $\Lambda$.

The exponent $\gamma_s$ determines the proliferation into so-called *baby universes* [7, 8] and the scaling exponent $\nu$ can be identified with the inverse Hausdorff dimension of the system:

$$\nu = 1/d_H. \tag{8}$$

We denote $d_H$ the *grand canonical (intrinsic) Hausdorff dimension*, since it is defined in the ensemble of manifolds with fluctuating volume $V_g = \int d^d\xi \sqrt{g}$, but constant cosmological constant $\Lambda$. Eq. (8) is reasonable since the average volume of the ensemble of manifolds with partition function $G_\Lambda(R)$ is:

$$\langle V_g \rangle_R = -\frac{\partial \ln G_\Lambda(R)}{\partial \Lambda} \sim R^{1/\nu}, \tag{9}$$

for $R \sim \Lambda^{-\nu}$. This relation follows from scaling behavior (5).

In the case of pure two-dimensional Euclidean quantum gravity one can calculate $G_\Lambda(R)$ analytically [1]:

$$G_\Lambda(R) = \Lambda^{3/4} \frac{\cosh \Lambda^{1/4} R}{\sinh^3 \Lambda^{1/4} R}. \tag{10}$$

If we expand $G_\Lambda(R)$ for small $R$ we get:

$$G_\Lambda(R) = \frac{1}{R^3} - \frac{1}{15}\Lambda R + \frac{4}{189}\Lambda^{3/2} R^3 + O(\Lambda^2 R^5), \tag{11}$$

i.e. the first term is indeed singular for small $R$ and analytic in $\Lambda$.

We can define the two-point function $G_V(R)$ on the ensemble of metric manifolds contributing to the partition function (2) in the same way as we defined $G_L(R)$ corresponding to the ensemble (1). It can be viewed as the partition function for universes with a fixed volume $V$ and two marked points separated by a geodesic distance $R$. $G_V(R)$ is related to $G_\Lambda(R)$ by a Laplace transformation, in the same way as $Z(\Lambda)$ is related to $Z(V)$ by (3):

$$G_\Lambda(R) = \int_0^\infty dV \, e^{-\Lambda V} \, G_V(R), \tag{12}$$

and $G_V(R)$ is by definition related to the average volume $S(R)dR$ in a spherical shell of thickness $dR$ a geodesic distance $R$ from a given point:

$$\langle S(R) \rangle_V = \frac{1}{V} \frac{G_V(R)}{Z(V)} = R^{d_h-1} \, F\left(\frac{R}{V^\nu}\right), \tag{13}$$

where $F(x)$ is a function with $F(0) > 0$, which falls off for large $x$. Eq. (13) is another definition of the (intrinsic) Hausdorff dimension associated with the *canonical ensemble* of metric manifolds used in two-dimensional Euclidean quantum gravity. We denote $d_h$ the *canonical Hausdorff dimension*. A priori there is no reason that $d_h = d_H$, but in fact it follows from the definitions that *if $F(0) > 0$ then $d_h = d_H$*. Clearly $F(0) > 0$ is a necessity for (13) to be valid as a definition of the Hausdorff dimension in the canonical ensemble. But from (5) it follows by inverse Laplace transformation and the assumption (13) with $F(0) > 0$ that

$$G_V(R) \sim V^{\gamma_s - 1 - \nu d_h} R^{d_h - 1} \quad \text{for} \quad R \to 0. \tag{14}$$

This is only compatible with (13) if $\nu d_h = 1$. In the following we will not distinguish the two Hausdorff dimensions, but completeness it should be mentioned that one can find non-unitary models where $d_h > d_H$. An example is the so called multicritical branched polymer model [9], where $d_h = 2$, while $d_H = m/(m-1)$, and $m \geq 3$ is an integer. From the relation between $G_\Lambda(R)$ and $G_V(R)$ we deduce that

$$G_\Lambda(R) \sim R^{\gamma_s/\nu - 1} \tilde{F}(\Lambda^\nu R), \qquad (15)$$

where $\tilde{F}$ is related to $F$ by

$$\tilde{F}(x) \sim \int_0^\infty dy e^{-yx^{d_h}} y^{\gamma - 2} F(\frac{1}{y^\nu}). \qquad (16)$$

We have that $\tilde{F}(0) > 0$ since the integral in (16) is well defined for $x = 0$, as $F(y)$ is expected to fall off faster than a power of $y$ (see (19) below) for $y \to \infty$, and $\gamma < 1$. In fact we can say somewhat more about $G_\Lambda(R)$ since the integral after $R$ should be as in eq. (7). This implies that we should be able to split $G_\Lambda(R)$ as follows:

$$G_\Lambda(R) = R^{\gamma_s/\nu - 1} P(\Lambda R^{d_h}) + \Lambda^{1-\gamma_s} R^{d_h - 1} \bar{F}(\Lambda R^{d_h}), \qquad (17)$$

where $P(x)$ is a polynomial of degree less than $[1 - \gamma_s]$ with $P(0) > 0$, while $\bar{F}(x)$ is a non-analytical function, also with $\bar{F}(0) > 0$. $P$ will not contribute to the singular term $\Lambda^{-\gamma_s}$ in (7) and the entire contribution comes from the integral over the last term in (17). Under an inverse Laplace transformation the first term in eq. (17) will not contribute to $G_V(R)$, $V > 0$, and in this way the small $R$ dependence $\Lambda^{1-\gamma_s} R^{d_h - 1}$ of the second term in (17) is the analogy of the term $V^{\gamma - 2} R^{d_h - 1}$ in $G_V(R)$. It is seen that this scenario is indeed satisfied in for the exact solution (10)-(11) of pure gravity. Further aspects of the expansion of $G_V(R)$ is discussed in the appendix.

From general arguments we finally expect that $G_\Lambda(R)$ falls off exponentially [4]

$$G_\Lambda(R) \sim e^{-\text{const.}\Lambda^\nu R} \qquad R \gg \Lambda^{-\nu} \qquad (18)$$

and this translates by inverse Laplace transformation to

$$G_V(R) \sim e^{-\text{const.}(R/V^\nu)^{1/(1-\nu)}} \qquad R \gg V^\nu. \qquad (19)$$

While (18) and (19) are exact in the case of pure gravity, they are unfortunately only bounds in the case where we have matter coupled to gravity.

A convenient quantity from a numerical point of view is the normalized distribution

$$n_V(R) = \frac{1}{V} \langle S(R) \rangle_V = V^{-\nu} x^{d_h - 1} F(x), \qquad x = \frac{R}{V^\nu}, \qquad (20)$$

such that

$$1 = \int_0^\infty dR \, n_V(R) = \int_0^\infty dx \, x^{d_h - 1} F(x). \qquad (21)$$

Eq. (20) has the form of a finite size scaling relation in the sense that it is a universal function of the reduced variable $x$ times a scale factor $V^{-\nu}$. In numerical simulations eqs. (20) and (21) turn out to be quite useful.

The Hausdorff dimension is not the only dimension which can naturally be defined on an ensemble of manifolds. The so-called spectral dimension is defined as follows: For a given manifold the propagation of a massless scalar particle is described by the inverse Laplacian, $\Delta_g^{-1}$, where

$$\Delta_g = \frac{1}{\sqrt{g}} \partial_a \sqrt{g} g^{ab} \partial_b. \tag{22}$$

The inverse Laplacian has a heat kernel representation,

$$\Delta_g^{-1}(\xi, \xi') = \int_0^\infty dT \left\langle \xi' | e^{T\Delta_g} | \xi \right\rangle, \tag{23}$$

where the heat kernel $K_g(\xi, \xi', T) = \left\langle \xi' | e^{T\Delta_g} | \xi \right\rangle$ is the kernel of the diffusion equation

$$\frac{\partial}{\partial T} \Phi = \Delta_g \Phi, \tag{24}$$

and the normalization is $\int d^d\xi \sqrt{g(\xi)} \, K_g(\xi, \xi', T) = 1$. In this paper we consider the initial condition,

$$K_g(\xi, \xi', 0) = \langle \xi' | \xi \rangle = \frac{1}{\sqrt{g(\xi)}} \delta(\xi - \xi'). \tag{25}$$

To the diffusion equation we can in the usual way associate a random walk interpretation. *In this way $K(\xi, \xi', T)$ is the probability per unit volume that a random walk, which starts at $\xi$ will be at $\xi'$ at "time" $T$.*

The heat kernel $K(\xi, \xi', T)$ has the asymptotic short distance expansion (the so-called Hadamard-De Witt-Minakshisundaram expansion, see for instance [10] for a review)

$$K_g(\xi, \xi', T) = \frac{e^{-d_g^2(\xi,\xi')/4T}}{(4\pi T)^{d/2}} \Lambda(\xi, \xi', T), \tag{26}$$

where $d_g(\xi, \xi')$ again denotes the geodesic distance and

$$\Lambda(\xi, \xi', T) \sim \sum_{r=0}^\infty a_r(\xi, \xi') T^r, \qquad \Lambda(\xi, \xi, 0) = 1. \tag{27}$$

The functions $a_r(\xi, \xi')$ satisfy certain invariant differential equations, and in the coincidence limit $\xi' \to \xi$ one can express $a_r(\xi, \xi)$ entirely in terms of local invariants, of which the first ones are:

$$\begin{aligned} a_0(\xi, \xi) &= 1, \\ a_1(\xi, \xi) &= \frac{1}{6}\mathcal{R}, \\ a_2(\xi, \xi) &= \frac{1}{72}\mathcal{R}^2 + \frac{1}{180}(\mathcal{R}_{abcd}\mathcal{R}^{abcd} - \mathcal{R}_{ab}\mathcal{R}^{ab}) + \frac{1}{30}\Delta_g \mathcal{R}, \end{aligned} \tag{28}$$

where $\mathcal{R}_{abcd}$, $\mathcal{R}_{ab}$, and $\mathcal{R}$ are curvature tensors. By taking the trace of the operator $\hat{K}_g$ corresponding to the kernel $K_g(\xi, \xi', T)$, we get

$$\operatorname{Tr} \hat{K}_g(T) \equiv \int d^d\xi \sqrt{g(\xi)} \, K_g(\xi, \xi, T) \sim \frac{1}{T^{d_s/2}} \sum_{r=0}^\infty A_r T^r, \tag{29}$$

where

$$A_r = \int d^d\xi \sqrt{g(\xi)} \, a_r(\xi, \xi). \tag{30}$$

We call the power $d_s$ the *spectral dimension*. For a smooth manifold we have $d_s = d$. However, it is possible to define diffusion and consequently the spectral dimension $d_s$ on more general structures than manifolds. We will discuss such cases later. If $V$ is the volume of the manifold it is seen that

$$\frac{1}{V}\text{Tr }\hat{K}_g(T) = \text{average return probability of a random walk at "time" } T. \qquad (31)$$

Since it is known that the kernel $K(\xi, \xi', T)$ can be expressed entirely in terms of reparametrization invariant quantities, as indicated in the asymptotic expansion (26), it makes sense to talk about the average in quantum gravity. For the asymptotic expansion (29) we get

$$\left\langle \text{Tr }\hat{K}_g(T) \right\rangle \sim \frac{1}{T^{d_s/2}} \sum_{r=0}^{\infty} \langle A_r \rangle T^r, \qquad (32)$$

where the quantum gravity average is taken over the operators which enter in (28)-(30). Strictly speaking we cannot be entirely sure that the quantum average can be taken term by term and we define the *quantum spectral dimension* $\bar{d}_s$ by

$$\left\langle \text{Tr }\hat{K}_g(T) \right\rangle \sim \frac{1}{T^{\bar{d}_s/2}} \quad \text{for} \quad T \to 0. \qquad (33)$$

and in case we consider the average over manifolds with fixed volumes the quantum return probability will be given by

$$\frac{1}{V} \left\langle \text{Tr }\hat{K}_g(T) \right\rangle_V. \qquad (34)$$

More generally we expect the behavior of (26) to be replaced by one which involves the geodesic distance as defined for the volume-volume correlator:

$$\bar{K}_V(R, T) \sim \frac{R^{d_h - 1}}{T^{\bar{d}_s/2}} H(R^{2d_h/\bar{d}_s}/T), \qquad R \ll V^{1/d_h} \qquad (35)$$

where $V$ is a fixed volume of the manifold, $H(0) > 0$ and[3]

$$\bar{K}_V(R, T) \equiv \frac{1}{VZ(V)} \iint \mathcal{D}[g_{ab}] \mathcal{D}\phi \, \delta(\int \sqrt{g} - V) e^{-S[g,\phi;G]}$$
$$\times \iint d^d\xi \sqrt{g(\xi)} d^d\xi' \sqrt{g(\xi')} \, K_g(\xi, \xi', T) \, \delta(d_g(\xi, \xi') - R). \qquad (36)$$

With this definition it follows that

$$\int_0^\infty dR \, \bar{K}_V(R, T) = 1, \quad \lim_{R \to 0} \frac{\bar{K}_V(R, T)}{\langle S(R) \rangle_V} = \frac{1}{V} \left\langle \text{Tr }\hat{K}_g(T) \right\rangle_V, \qquad (37)$$

and that the average geodesic distance travel by diffusion over time $T$ is given by:

$$\langle R(T) \rangle_V = \int_0^\infty dR \, R \, \bar{K}_V(R, T) \sim T^{\bar{d}_s/2d_h}. \qquad (38)$$

It is easy to understand that $d_h$ can be different from $d$ in quantum gravity. We defined $d_h$ by

$$\lim_{R \to 0} \langle S(R) \rangle_V \sim R^{d_h - 1}, \qquad (39)$$

---

[3] An alternative definition would be one where we divide with $G_V(R)$ in (36), rather than with $VZ(V)$. as in (36). By (13) it corresponds to the removal of the factor $R^{d_h - 1}$ from $\bar{K}_V(R, T)$.

where the average is performed over all metric manifolds of fixed topology and volume $V$. If we could take the limit $R \to 0$ for each (smooth) manifold, before taking the quantum average, we would get $d_h = d$. However, there is no reason such an interchange of limits should be possible, and as mentioned above, it is not possible for pure gravity in $d = 2$, where $d_h = 4$. For the spectral dimension the situation is less clear. It seems quite reasonable that the asymptotic expansion (32) is valid, in which case the spectral dimension is $\bar{d}_s = d$. However, even in this case we will find some situations where the spectral dimension $\bar{d}_s$ is different from $d$.

The purpose of this article is to report on extensive numerical simulations where we determine the Hausdorff dimension as well as the quantum spectral dimension of quantum gravity coupled to various matter fields[4].

The rest of this article is organized as follows. In sec. 2 we outline the numerical setup and in sec. 3 we test the scaling prediction (20) for pure gravity, while sec. 4 contains the generalization to the theory of gravity coupled to matter fields. Sec. 5 deals with the analysis of the so-called spectral dimension. In sec. 6 we analyse the use of finite size scaling relations in quantum gravity and make predictions about the behavior of reparametrization invariant two-point functions. Finally, sec. 7 contains a critical discussion of the results obtained so far.

## 2 The numerical method

The numerical setup of the simulations presented in this paper is as follows: For all studied systems we use the discretization commonly known under the name of the "dynamical triangulation". We consider ensembles of surfaces built of equilateral triangles with spherical topology and a fixed number of triangles. These surfaces can be viewed as dual to planar $\phi^3$ diagrams. In the language of the $\phi^3$ theory we include diagrams containing tadpole and self–energy subdiagrams. In the language of the direct lattice it means that we allow "triangulations" where two vertices are joined by an arbitrary number of links and where a link can join a vertex to itself. The coordination number of a point can be any positive integer. All these diagrams are present in the simplest matrix model formulation, without the necessity to renormalize the couplings. In earlier works it has been noted that inclusion of these subdiagrams actually speeds up the convergence when critical indices are measured (cf. [11]). A similar formulation was used in [5]. In case matter fields were used in the simulations their interactions were such that they could be viewed as placed in the centers of triangles (or on vertices of the dual $\phi^3$ graphs).

A new element in simulation is the use of so called generalized *minbu surgery*[5] in the update scheme of the surface geometry. These are global rearrangments of the triangulation which are used in addition to the the standard "flips". They are closely related to the "minbu surgery moves" described in [12] for the two–dimensional system *without* tadpole and self–energy subdiagrams or in [2] for the four–dimensional systems. For the sake of the present simulation we define a "minbu" to be a smaller part of the triangulation separated from the remaining part by two links, joining the same two vertices (in the dual $\phi^3$ language this corresponds to a special form of the self–energy subdiagram, special in the sense that we exclude the case when two vertices have the same label). The two links form a "neck" of the minbu. The new algorithm consists of two steps: in the first the surface is cut along the neck and the boundaries are closed. In effect the spherical surface splits into two surfaces both with spherical topology. In the second step we select at random a link on each surface joining two points with different labels. The surfaces are cut along these links and glued together along the boundaries, the chosen links becoming a neck of the new minbu. It is rather simple to work out the detailed balance condition for this type of move. In practice we put a lower limit on the size of the minbu to avoid moving numerous small minbus.

---

[4]Closely related work has recently appeared in [5]. Where overlapping, the conclusions are indentical. However, as we will show, the situation in pure gravity is actually much better than it appears from the "raw" data presented in [5].

[5]*Minbu* is an abbreviation for "miminal neck baby universe" [7, 8].

Both the standard flips and the new global moves are organized in "sweeps". In case of the flips they correspond to the number of attempted flips being equal the number of links. For the minbu surgery moves a sweep corresponds to the number of attempted moves of this type being equal to the number of minbus on the surface before the sweep (this number usually changes during the sweep). A global sweep was in all cases followed by 9 standard sweeps (the global moves alone do not satisfy the ergodicity requirement). The new move helps to reduce the correlation times and it's effect is quite dramatic (cf. [12] and [2] for discussion of this point). It also gives for free a possibility to measure $\gamma_{str}$ (cf. [7, 8]). The results presented in this paper corespond to system sizes 1000, 2000, 4000, 8000, 16000 and 32000 triangles. For these systems the lower limits on the minbu size were chosen in the range from 10 to 20 triangles. Since the update of the system is "cheap" in computer time, the typical measurements were performed every 200 sweeps, safely above the longest autocorrelation times observed and the number of measurements in a typical experiment ranged between 1000 and 5000. In the analysis of the diffusion equation only triangulation with 4000 and 16000 triangles were used because of the large measurement times. In this case the measurements form the real time barrier, taking up to 95% of the computer time for the larger systems and we were forced to reduce the number of measurements. A typical experiment in this case was an analysis based on 100 configurations, separated by 1000 sweeps.

## 3 Pure gravity

When we use the formalism of dynamical triangulations as a regularization of the theory of quantum gravity the discretized volume is identified with the number of triangles $N$:

$$V = N\, a^2, \qquad (40)$$

where $a^2$ is the area of each triangle. In addition we identify the volume elements $d^2\xi\sqrt{g}$ with the area of the triangles and the geodesic distance $r$ between two triangles as the shortest path along neighboring triangles. With this definition the discretized geodesic distance is always an integer. This definition is only one among many possible. We are later going to test the results for other definitions of the geodesic distance. In the scaling limit they should all be equivalent. If this is not the case, it is difficult to have any confidence in the "continuum" results extracted from the discretized theory.

Pure two-dimensional gravity is a good test case for numerical simulations since we know the exact formula for $G_\Lambda(R)$ and consequently for $x^{d_h-1}F(x)$. In the appendix we outline how to find $x^{d_h-1}F(x)$, which turned out to be sligtly non-trivial.

We denote the discretized version of $n_V(R)$ by $n_N(r)$. It satisfies

$$\sum_{r=0}^{N} n_N(r) = 1. \qquad (41)$$

It is easy to obtain $n_N(r)$ directly from the measurements and we can check if the data are compatible with the form (20), i.e. if it is possible to find a number $d_h$ such that

$$N^{1/d_h} n_N(r) = x^{d_h-1} F(x), \qquad x = r/N^{1/d_h}, \tag{42}$$

for various sizes $N$ of the system. In principle there is an additional constant of proportionality $\alpha$ between $x$ and $r$, if $x$ refers to the continuum variable in (20). This constant depends on the regularization (e.g. which class of triangulations one uses in the discretization, the definition of geodesic distance etc). In the appendix we give this constant for the class of triangulations used in our computer simulations and our definition of geodesic distance. In (42) and the rest of the article this constant has been absorbed in a redefinition of $x$ in order not to make the notation unnecessary cumbersome.

While it is possible to find a reasonable value of $d_h$ this way, a much better result is obtained by performing a shift in the values of $r$ and $N$ before applying the scaling relation (42). This is reasonable from the point of view that the shortest distances and volumes are lattice artifacts where we can not expect agreement with the continuum formulas. In this way we are led to a "phenomenological" scaling variable $x$

$$x = \frac{r+a}{N^{1/d_h} + b}. \tag{43}$$

We can "derive" (43) from the data in the following way: fix $d_h = 4$ and fix $N$ and determine for a given value of $r$ the value of $x$ such that the lhs of (42) (the measured function) agrees with the rhs (the calculated continuum function). In fig. 1 we have shown $r$ as function of $x$ for different volumes $N$. We observe a perfect linear relation except for the smallest values of $r$ and the constant $a$ is approximately independent of $N$. The slope changes with $N$ and we can fit it well to $(N^{1/4} + b)$. This is shown in fig. 2, where the result of a common fit like (43) to the data involving $N = 1000, 2000, 4000, 8000, 16000$ and $32000$ triangles are shown with the factor $N^{1/4} + b$ divided out. All data are contained in this graph and we see that the phenomenological scaling (43) is well satisfied.

One important lesson from fig. 1 is that a relation like (43) is not valid for the smallest values of $r$ and one should simply discard these small values.

In fig. 3 we have shown the data and the theoretical curve for $d_h = 4$ and an optimal choice of $a$ and $b$. The agreement is almost perfect and the conclusion is that *we already see continuum physics for systems as small as 1000 triangles in the case of pure gravity* if we include simple finite size corrections like (43) A different graphical representation of the data, well suited for the small $x$ region, is obtained by plotting $d \log n_N(r(x))/d \log x$ versus $x$. It has the virtue that it tests clearly the consistency of $d_H = d_h$, since the value $d_H$ which provides us with overlapping graphs should agree with the asymptotic value of the curve for small $x$, which determines $d_h$. We have illustrated this in fig. 4, where the theoretical curve is also displayed.

As mentioned above there are many different ways to define the concept of geodesic distance on the ensemble of piecewise linear manifolds. One alternative definition is as the shortest "link" distance between two vertices. To each vertex $v$ we assign an area element

$$dA_v = \frac{1}{3} n_v a^2, \tag{44}$$

where $n_v$ is the order of the vertex. $dA_v$ is assumed to replace the continuum $d^2 \xi \sqrt{g}$ when we form the average. For a given triangulation the "link-distance" and the "triangle-distance" used above (the link distance on the dual $\phi^3$-graph) can differ vastly. However, as ensemble average we expect that they agree up to some constant of proportionality. This was proven analytically for pure two-dimensional gravity in [1]. Numerically it is also seen very clearly.

In the next section we will study the Hausdorff dimension when matter is coupled to gravity. In these cases we do not know the Hausdorff dimension. Following the philosophy outlined above one should now determine the constants $a$, $b$ and $d_h$ such that the distributions $n_N(r)$ can be mapped into each other as a function of $x$. The best $d_h$ determined this way from (43) would then be the candidate for a Hausdorff dimension. If we do this, we get

$$d_h = 3.9 \pm 0.2. \qquad (45)$$

We view this result as the typical accuracy one can expect. For choices of $d_h$ in this interval we get good overlaps of the distributions with acceptable $\chi^2$. In order to get a more precise determination of $d_h$ one would have to improve the statistics of the large-$r$ tail of the distribution $n(r)$, which in principle is possible, but in practice is very computer intensive since the probability of creating these elongated universes is exponentially small.

One can use simpler charactistics of the distribution $n_N(r)$, like the fact that the peak of the distributions should scale as $N^{-1/d_h}$, to determine $d_h$. It gives results comparable to (45). However, we regard the procedure outlined above as more concicing, since the requirement that one can map the various distributions onto each other for different volumes is a much stronger test of scaling.

## 4 Matter fields

We can now perform the analysis outlined above in the case of matter coupled to gravity. We have performed extensive computer simulations for Ising spins coupled to gravity, three-state Potts model coupled to gravity and one to five Gaussian fields coupled to gravity. In the critical point the Ising model describes a $c = 1/2$ conformal field theory, the three-states Potts model a $c = 4/5$ conformal field theory[6], while the Gaussian fields automatically are critical with a central charge equal to the number of Gaussian fields.

For all these theories we can measure $n_N(r)$ and try to determine a possible $d_h$. For the theories considered so far, it seems that we *have* scaling according to (42) and (43). We have illustrated this for $c = 1/2$, 1, 2 and c=5 in fig.5. Each of the graphs contain the scaled data for system sizes 1000, 2000, 4000, 8000, 16000 and 32000 triangles. The results for other $c$'s ($c = 4/5$, 3 and 4) are similar. We have shown both the distributions $n(r(x))$, as well as the derivative distributions $d\log n(r(x))/d\log x$ which, as mentioned above, have the advantage of displaying the small $x$ region more clearly.

In these cases the Hausdorff dimensions have been *chosen* to be 4 for $c \leq 1$, 3 for $c = 2$ and 2 for $c = 5$, respectively, and we see as good scaling as in the case of pure gravity. As already remarked the internal consistency of the fitting to *a single Hausdorff dimension* (i.e $d_H = d_h$) requires that the logarithmic derivatives should converge to the chosen $d_h$ for small $x$. As is seen on the the plots on the left hand side of fig. 5 the graphs for $c = 1/2$ and $c = 1$ indicate a slightly lower $d_h$ (around 3.8). The same value was found to give the best scaling in for pure gravity, where it is known that $d_h = 4.0$. Also for the other values of $c \leq 1$ we find the best all over scaling for $d_h \approx 3.8$. Finally it should be mentioned that for $c \leq 1$ the best values of the constants $a$ and $b$ in the phenomenological formula (43)

---

[6]The value of the critical point of the three-states Potts model coupled to gravity has recently been calculated [13].

are very close to the pure gravity values. Clearly wee have some interval of $d_h$ where the fits appear to be acceptable and in this way we extract a tentative Hausdorff dimension for the various theories. The result of the best $d_h$ as well as errorbars of a somewhat subjective nature is shown in fig. 6. From the figure and the fact that $d = 4$ gives very good fits for $c \leq 1$ and $d_h = 2$ very good fits for $c = 4$ and 5, it is tempting to make the following

**Conjecture:** $d_h = 4$ for $c \leq 1$. $d_h = 2$ for large $c$.

This conjecture is in agreement with old numerical results from $c = -2$ systems where a measurement of the Hausdorff dimension gave results very close to $d_h = 4$. It is also in accordance with the recent numerical simulations reported in [5] for $c < 1$.

It may be surprising (but see sec. 7) that $d_h = 4$ for all $c \leq 1$, since there are indications of different scaling in pure gravity and gravity coupled to $(m, m+1)$ conformal matter. Using a Hamiltonian formulation of string field theory for non-critical strings it has been argued that the scaling should be governed by the dimensionless variable $\Lambda^{1/2m}T$, where $T$ is the so-called proper "time" [6]. This should be compared to the known result for pure gravity as presented in eq. (18). If we could identify $T$ with the geodesic distance one could conclude that $d_h = 2m$ for the $(m, m+1)$ model coupled to gravity. However, for $m > 2$ we have no such identification. It is in principle possible to have an identification $T = R^{2/m}$, in which case the conjectures from the Hamiltonian formalism would be in accordance with the observed $d_h = 4$. It is natural in the Hamiltonian formalism to expect[7] an exponential decay like (18) of the volume-volume correlator $G_\Lambda(R)$ with respect to $T$. Such a decay and $d_h = 4$ would lead to $G_\Lambda(R) \sim e^{-(\Lambda^{1/4}R)^{2/m}}$, i.e. a decay of $G_\Lambda(R)$ which is slower than exponentially, but not power like. Although unconventional, such behavior cannot be ruled out a priori, since it does not violate the bound (18). By inverse Laplace transformation we get (see also the second part of the appendix for further discussion)

$$G_V(R) \sim e^{-c(R/V^{\frac{1}{4}})^{4/(2m-1)}}. \tag{46}$$

This behavior is different from the one of pure gravity (see (19)) and can in principle be tested from the observed distribution $n_N(r)$ of the discretized theory since the prediction is:

$$\log n_N(r) \sim x^{4/(2m-1)} \quad \text{for} \quad x \gg 1, \tag{47}$$

where $x$ is the scaling variable in (42)-(43). Unfortunately this difference in behavior is for large $x$ and the statistics of the tail of the distribution $n_N(r)$ is not sufficiently good to distinguish between the power 4/3 for pure gravity and the power 4/5 corresponding to $m = 3$ for the Ising model. The reason is that we have to allow for unknown *subleading*, i.e. power like correction factors to (46) and that we have to consider quite large $x$ before we can ignore such factors. It would be most interesting if one could prove or disprove (47). The statistics is much better for small $x$ and as discussed in the appendix we expect the following behavior in the case of pure gravity:

$$n_N(r) \sim N^{-1/d_h} x^3 \left(1 + c_1 x^4 + c_2 x^8 + \cdots\right). \tag{48}$$

While $d_h = 4$ leads to the use of the scaling variable $x$ as well as to the leading term in (48), there is no reason the next terms should agree in pure gravity and after coupling to matter

---
[7]But we should stress that it has not yet been proven.

even if $d_h = 4$. In the appendix we present arguments which indicate that the leading correction term $x^4$ in (48) is replaced by an $x^{-\gamma_s/\nu}$ term in the case of an minimal unitary model coupled to gravity. This power is two or less in the case of the minimal models. We test the subleading corrections to $n_N(r(x))$ by plotting the logarithmic derivative as a function of $x^\alpha$:

$$\frac{d \log n(r(x))}{d \log x} = d_h - 1 + \alpha c_1 x^\alpha + o(x^\alpha), \qquad (49)$$

if $n(r(x)) \sim x^{d_h-1}(1 + c_1 x^\alpha + o(x^\alpha))$. The best power of $\alpha$ is the one where the data have a definite slope for $x$ small. In fig. 7 is shows the correction term for central charge $c = 0$, $c = 1/2$, $c = 1$ and $c = 5$. As long as $c \leq 1$ the results agree quite well with pure gravity, and the power $\alpha$ is definitely between 3 and 5, i.e. a very small deviation, if any, from the result of pure gravity. In order to explain this selection of powers of the subleading term it is tempting to

**Conjecture:** $G_V(R) \to e^{2\pi i \gamma_s} G_V(R)$  for  $V \to e^{2\pi i} V$  and  $0 \leq c < 1$.

It would be interesting to understand the reason for this "symmetry principle".

Up to this point we have not presented any evidence that coupling of matter to gravity creates a back-reaction on the geometry associated to the metric properties as long as $c \leq 1$. This is in marked contrast to the situation for $c \geq 2$. Here the Hausdorff dimension *did change* (cf. fig. 6) and fig. 7 shows that the subleading small $x$ corrections are different as well. The exponent $\alpha \approx 2$ for $c = 4$ and 5. However, even for $c \leq 1$ the distributions, scaled and shifted according to (42) and (43), *are not identical*, as shown in fig. 8. Note that the distributions agree very well for small $x$, in accordance with the hypothesis that $d_h = 4$ and that even the subleading exponents $\alpha$ in (49) are identical. Unfortunately we have not found a convincing parametrization of the difference between the various curves for $c \leq 1$ which is actually observed for $x > 3$. Only for large $x$ can we use a parametrization like (47) due to the problems with unknown subleading corrections, as mentioned above.

As mentioned in sec. 2 the string susceptibility is an aspect of the fractal structure of quantum gravity we get for free in our computer simulations if we use the "minbu surgery" update algorithm. In contrast to the Hausdorff dimension the string susceptibility $\gamma_s$ shows a clear dependence on the conformal matter coupled to gravity even for $c < 1$. In fig. 9 we have shown a measurement of $\gamma_s$ for the various matter theories. It should be compared to fig. 6. It is seen that $\gamma_s \to 1/2$ relatively fast above $c = 1$. All this corroborates on the idea that the scaling limit for $c$ large is that of branched polymers. They have Hausdorff dimension 2 and $\gamma = 1/2$. Several remarks are in order. While the measured $\gamma_s$ for $c < 1$ agree well with the theoretical results, $\gamma_s$ comes out too small for $c = 1$. This is a well known effect (cf. [14]) and is due to logarithmic corrections which are large. If they are included one gets $\gamma_s \approx 0$. These corrections have not been included in fig. 42 where we have preferred to treat all data set identical, i.e. in this case without logarithmic correction, which are not present for $c < 1$ and probably not for $c > 1$ either. The results for $c > 1$ are somewhat larger than the previously reported results [14]. However, the present measurements are performed with an ensemble of manifolds where the triangulations (represented as by their dual $\phi^3$ graphs) include tadpoles and self energy diagrams. This may explain the difference. It is somewhat remarkable that the values of $\gamma_s > 0$ seem consistent which a theorem [15] which states that if $\gamma_s > 0$ and all manifolds (after integrating over matter fields) are counted with positive weight, then

$\gamma_s = 1/m$, where $m \geq 2$.

# 5 The spectral dimension

We now turn to the measurement of the spectral dimension in two-dimensional quantum gravity.

We can measure the spectral dimension as defined by (33). At the discretized level it is natural to consider random paths between neighboring triangles. In this case the discretized diffusion equation can be written:

$$\phi(i, t+1) = \frac{1}{3} \sum_{(ij)} \phi(j, t), \tag{50}$$

where $(ij)$ is one precisely for the three triangles $j$ with are adjacent to triangle $i$, and zero elsewhere. The continuum normalization of $\Phi$ which corresponds to the heat kernel is $\Phi(\xi, t) = K(\xi, 0, t)$, and it translates to the discretized notation as follows:

$$\Phi_0(\xi, 0) = \frac{1}{\sqrt{g(\xi)}} \delta(\xi - \xi_0) \quad \rightarrow \quad \phi_0(i, 0) = \delta_{i, i_0}. \tag{51}$$

The measurement of the quantum spectral dimension as defined by (33) is performed by generating a number of independent triangulations by the standard Monte Carlo technique, and for each triangulation solve the diffusion equation (51) and measure the return probability $\Phi(i_0, t)$. The triangle $i_0$ should be picked with even probability since it represents unit area. From a practical point of view it is convenient to perform an average over some $i_0$ for each independent triangulation generated by the computer simulations.

As in the case of the Hausdorff dimension it is important to test if the results are independent of the detailed "microscopic" definition of the geodesic distance, since it is by no means universal. Again a simple test of the universality is obtained by studying the diffusion using the links instead of the triangles as possible paths, and the shortest link path between two vertices as a definition of the geodesic distance. As in (44) the area element associated with a vertex $v$ is proportional to the order $n_v$ of the vertex. The analogy of (51) will be

$$\Phi_0(\xi, 0) = \frac{1}{\sqrt{g(\xi)}} \delta(\xi - \xi_0) \quad \rightarrow \quad \phi_0(v, 0) = \frac{1}{n_v} \delta_{v, v_0}, \tag{52}$$

and the diffusion equation (50) will be replaced by:

$$\phi(v, t+1) = \frac{1}{n_v} \sum_{(vv')} \phi(v', t), \tag{53}$$

where the summation is over the $n_v$ neighboring vertices $v'$ to $v$. The results obtained this way essentially agree with the results obtained from (50) and (51) but are in fact considerable better behaved for small $t$ as discussed below. In the following we will show only the results coming from (52) and (53).

When comparing the measured return probability with the expected $t^{-\bar{d}_s/2}$ two points are important. The constant function is a normalizable solution of the diffusion equation on a compact manifold, and the behavior $t^{-\bar{d}_s/2}$ can only be valid for sufficiently small times on compact manifolds. For large times $\phi(i_0, t)$ will just be constant. Substracting the constant does not really help us, since the spectrum of the Laplacian is discrete on compact manifolds, and this just means that the next lowest eigenfunction will dominate for large $t$. We expect the behavior to be the correct one only up to times which are of order or the inverse of the lowest eigenvalue different from zero. The second problem is that the behavior $t^{-\bar{d}_s/2}$ is not correct for small $t$ either, due to the discretization. Clearly it makes no sense to talk about $t < 1$, but the situation is slightly worse, as is illustrated by considering a discretized random walk in one dimension. In this case the return probability is identical zero for odd discretized times $t$! Here we observe for small $t$ a marked asymmetry between odd and even $t$, and it is considerable more pronounced and last for longer time if we use the return probability $\phi(i_0, t)$ coming from (50)-(51) rather than if we use $\phi(v_0, t)$ from (52)-(53).

In principle these problems should disappear in the continuum limit, since the eigenvalue density increases with increasing discretized volume[8] $N$. However, in order to use as efficiently as possible the small $t$ behavior we have imitated a *continuous* time in the following way: We can write the solution to (53) as follows

$$\phi(v, t) = (1 + \hat{\Delta}^{\text{dscr}})^t_{vv'} \phi_0(v', 0), \tag{54}$$

where the $V \times V$ *matrix* $\hat{\Delta}^{\text{dscr}}_{vv'}$ ($V$ being the number of vertices in the triangulation) is the discrete Laplacian corresponding to the diffusion equation (53)

$$\hat{\Delta}^{\text{dscr}}_{vv'} \phi(v') = \frac{1}{n_v} \sum_{(vv')} (\phi(v') - \phi(v)) \tag{55}$$

where the summation again is over the vertices $v'$ which are neighboring $v$. The solution to the continuous diffusion equation (24) is given by

$$\Phi(\xi, t) = \int d^d \xi' \left[ e^{T \hat{\Delta}_g} \right] (\xi, \xi') \Phi_0(\xi', 0). \tag{56}$$

We copy this formula by replacing the solution (54) by

$$\tilde{\phi}(v, t) = (1 + \frac{t}{n} \hat{\Delta}^{\text{dscr}})^n_{vv'} \phi_0(v', 0), \tag{57}$$

where $n > t$. For $n \to \infty$ the operator in (57) obviously goes $e^{t \hat{\Delta}^{\text{dscr}}}$ which is what we want. From a practical point of view it has the advantage that simply evolving the original diffusion equation (53) $n$ steps we know the vectors $[(1 + \hat{\Delta}^{\text{dscr}})^n \phi_0](v)$, and this is all we need in order to calculate $\tilde{\phi}(v, t)$ for $t < n$. This imitation of continuous time behavior (but with a discrete Laplacian $\hat{\Delta}^{\text{dscr}}$) yields a much smoother small $t$ behavior, and the data in fig. 10 and fig. 11 are obtained this way.

In fig. 10 we have shown the typical return probability as a function of time for $c = 0$ (pure gravity), $c = 1/2$, $c = 1$ and $c = 2, 3, 4, 5$. *We observe that the spectral dimension is consistent with 2 for $c \leq 1$ and that it decreases for $c > 1$.*

Since we solve the diffusion equation on each of the generated manifolds we can in addition determine the average length of diffusion at time $t$ and compare with the theoretical formula (38). The results are shown in fig. 11 for $c$ =0, 1, 3 and 5, and they are consistent with a dependence

$$\langle r \rangle_t \sim t^4. \tag{58}$$

---

[8]While the eigenvalue density increases at the discretized level, the continuum spectrum should be (approximately) constant (and of course discrete) since the continuum volume $V = Na^2$, $a$ being the lattice spacing, is assumed to be constant. It comes about because we have to multiply each discrete eigenvalue of the discretized Laplacian with $1/a^2$ in order to approximate the continuum Laplacian.

This lead us to

**Conjecture:** $\bar{d}_s = 2$ for the central charge $c \leq 1$. $d_h = 2\bar{d}_s$ for all finite values of $c$.

# 6  Finite size scaling

Finite size scaling has been a very important tool in the analysis of critical systems in statistical mechanics. Likewise it has been a convenient tool in the analysis of computer simulations of quantum gravity [16]. However, in the latter case the theoretical basis is not well understood. It is the purpose of this section to present the present situation in two-dimensional quantum gravity, where we have available some analytical tools.

One basic assumption in the theory of critical phenomena is that of a divergent correlation length. Let us consider a $d$-dimensional spin system where the inverse temperature is denoted $\beta$, the inverse critical temperature $\beta_c$, and

$$t \equiv (\beta_c - \beta)/\beta_c. \tag{59}$$

We denote the correlation length which diverges for $t \to 0$ by $\xi(t)$:

$$\xi(t) \sim t^{-\nu} \quad \text{for} \quad t \to 0. \tag{60}$$

The *scaling hypothesis* states that $\xi$ is the only relevant intrinsic scale. Assume that the singular part of the free energy per unit volume $f(\beta)$ behaves like $t^{2-\alpha}$. If we are slightly above the critical temperature, a fluctuation away from the ordered state by $\xi$ produces an increase $\Delta f \sim t^{2-\alpha}\xi^d$. The probability of such fluctuation is $e^{-\Delta f}$, and it becomes small for $\Delta f > 1$. By this heuristic argument one arrives at $\xi(t) \sim t^{-(2-\alpha)/d}$, i.e.:

$$f(t) = t^{\nu d} \quad \text{and} \quad \alpha = 2 - \nu d. \tag{61}$$

Assuming (61) is called the *hyperscaling hypothesis*. $\alpha$ is the exponent of the singular part of the specific heat $c(\beta)$, since we get the specific heat by differentiating $f(\beta)$ two times after $\beta$. Differentiating one time we get the singular part of the internal energy $\varepsilon(t)$, i.e.

$$\varepsilon(t) \sim t^{1-\alpha}, \qquad c(t) \sim t^{-\alpha}. \tag{62}$$

The spin-spin fluctuations will be long ranged when $t \to 0$. From general arguments one expects the following behavior:

$$\langle \phi(r)\phi(0) \rangle \sim \frac{1}{r^{d-2+\eta}} \, g(\frac{r}{\xi(t)}), \tag{63}$$

where $g(0) = 1$ and $g(x)$ falls off exponentially for large $x$. The second derivative of the free energy per unit volume with respect to an external magnetic field is the magnetic (or spin) susceptibility

$$\chi(t) = \int d^d x \, \langle \phi(x)\phi(0) \rangle \sim \xi(t)^{2-\eta}, \tag{64}$$

where $\phi(x)$ denote the continuum limit of the spin field, i.e. we have

$$\chi(t) \sim t^{-\gamma_m}, \quad \text{and} \quad \gamma_m = \nu(2-\eta), \tag{65}$$

where $\gamma_m$ is the magnetic susceptibility. These standard scaling relations can partly be derived by renormalization group arguments and they are readily converted into finite size scaling relations by assuming that the peak in specific heat and spin susceptibility occur at the so-called pseudo critical point where

$$t^{\nu d} \sim V^{-1}, \tag{66}$$

$V$ denoting the bulk volume of the system. It just tells us that the correlation length is of the order of the linear extension of the system and that singular part of the free energy $F(t) = Vf(t)$ is of order one at this point. By substituting (66) in the formulas above we get that the singular parts of the specific heat and spin susceptibility at the pseudo critical point behave as:

$$\chi(V) \sim V^{\gamma_m/\nu d}, \quad c(V) \sim V^{\alpha/\nu d}. \tag{67}$$

The set of finite size scaling relations is most readily derived by assuming that certain fields and composite operators $\Phi_i$, here the spin field and the energy operator, have well-defined scaling dimensions

$$\Delta_i^{(0)} = (d-2+\eta_i)/2 \tag{68}$$

close to the fixed point. If this is the case, we get for dimensional reasons:

$$\left\langle \int_V d^d x_1 \cdots \int_V d^d x_n \Phi_1(x_1) \cdots \Phi_n(x_n) \right\rangle \sim V^{n-(\Delta_1^{(0)}+\cdots \Delta_n^{(0)})/d}. \tag{69}$$

In the case of the spin system this allows us to express the critical exponents in terms of the scaling dimensions of the spin field $\Delta_\sigma^{(0)}$ and the energy density $\Delta_\varepsilon^{(0)}$. In the case of the energy density we get $\varepsilon(V) \sim V^{-\Delta_\varepsilon^{(0)}/d}$, while for the magnetization we get $m(V) \sim V^{-\Delta_\sigma^{(0)}/d}$. A factor $1/V$ has been divided out compared to (69), since the quantities $\varepsilon(V)$ and $m(V)$ are defined per unit volume. For the specific heat and the spin susceptibility we get:

$$c(V) = \frac{1}{V} \int_V d^d x \int_V d^d y \, \langle \Phi_\varepsilon(x) \Phi_\varepsilon(y) \rangle \sim V^{1-2\Delta_\varepsilon^{(0)}/d}, \tag{70}$$

$$\chi(V) = \frac{1}{V} \int_V d^d x \int_V d^d y \, \langle \Phi_\sigma(x) \Phi_\sigma(y) \rangle \sim V^{1-2\Delta_\sigma^{(0)}/d}. \tag{71}$$

If we compare with (67) we can write:

$$\alpha = \nu d(1 - 2\Delta_\varepsilon^{(0)}/d), \quad \gamma_m = \nu d(1 - 2\Delta_\sigma^{(0)}/d). \tag{72}$$

Finally the relation of the $\Delta_\varepsilon^{(0)}$ and $\Delta_\sigma^{(0)}$ to the renormalization group is established by the well-known fact that the *relevant* eigenvalues under a scaling with $\lambda$ are given by $\lambda^{1/\nu}$ and $\lambda^{y_h}$ where

$$\frac{1}{\nu} = d - \Delta_\varepsilon^{(0)}, \quad y_h = d - \Delta_\sigma^{(0)}. \tag{73}$$

The scaling ansatz of DDK [17] generalizes (70)-(71) to 2d quantum gravity by simply replacing $\int_V d^2 x$ by $\int d^2 x \sqrt{g}$, moving the ket and bra outside the integration in order to include the gravitational average over $2d$-manifolds with volume $V$. For a conformal field theory with central charge $c$ the change in $\Delta^{(0)}$ is given by:

$$\Delta = \frac{\sqrt{25-c+12\Delta^{(0)}} - \sqrt{1-c}}{\sqrt{25-c} - \sqrt{1-c}}. \tag{74}$$

This is the finite size scaling according to DDK and in this way it is possible to calculated the gravity modified exponents $\alpha$, $\beta$ and $\gamma_m$ for the spin system as long as $c \leq 1$. What is missing in order to get a description as detailed as that of ordinary statistical mechanics is the concept of a divergent correlation length and a corresponding relation to the renormalization group. The concept of geodesic distance, as introduced in the last section, provides a natural framework in which it should be possible to have relations like (63) and (64). A possible definition of the two-point function for a given field $\phi$ and fixed volume $V$ is[9]

$$\langle \phi(r)\phi(0) \rangle_V = \frac{1}{VZ(V)} \iint \mathcal{D}[g_{ab}]\mathcal{D}\phi\, e^{-S} \iint \sqrt{g(\xi)}\sqrt{g(\xi')}\, \phi(\xi)\phi(\xi')\, \delta(d(\xi,\xi') - r), \tag{75}$$

where $S$ denotes the combined action of gravity and matter fields and the $Z(V)$ denotes the partition function for finite volume $V$.

From dimensional arguments we have to conjecture that the quantum gravity generalization of (63) and (68) should be

$$\langle \phi(r)\phi(0) \rangle_V \sim \frac{r^{d_h-1}}{r^{2\Delta_\phi d_h/d}} g\left(\frac{r}{\xi(t)}\right), \tag{76}$$

where $d_h$ is the Hausdorff dimension and $\xi(t)$ the correlation length, *measured in geodesic distance unit*. For $t \to 0$ we assume that $\xi(t)$ diverges for infinite $V$, while $f(0)$ is finite. This conjecture leads to the correct scaling

$$\int_0^{V^{1/d_h}} dr\, \langle \phi(r)\phi(0) \rangle_V \sim V^{1-2\Delta_\phi/d} \tag{77}$$

at the critical point. If we assume that $\xi(t)$ divergences as $t^{-\nu}$ we get in addition the analogue of Fishers scaling relation:

$$\gamma_\phi = \nu d_h (1 - 2\Delta_\phi/d), \tag{78}$$

where $\gamma_\phi$ is the susceptibility exponent for the $\phi$-$\phi$ correlator, defined as for the ordinary spin-spin correlator. It should be compared to (72). $d$ has been replaced by $d_h$ when combined with $\nu$ and the "bare" scaling exponent by the dressed one, related as in (74) in the two-dimensional case.

The conjecture (76) has the appealing feature that it attributes a physical interpretation of the scaling dimension at the level of correlation functions, precisely as for a fixed manifold, *but the question is whether it is correct*. It has not yet been proven that there exits a divergent correlation length $\xi(t)$ in quantum gravity in the way defined above. We view this as one of the most interesting unsolved problems in two-dimensional quantum gravity. In principle the answer can be found by numerical simulations, although it seems difficult presently to measure the spin-spin correlation functions with sufficiently accuracy[10]. The Ising model coupled to quantum gravity seems to be an ideal test model for the conjecture (76).

---

[9]It is possible to adapt several, slightly different definitions of the correlation functions in quantum gravity. The one chosen here includes an angular average. We could choose to divide by this angular average, either inside the functional integration in (75), or outside the the functional integration, in which case one (by (13)) should replace the normalization $VZ(V)$ by $G_V(R)$. In addition one should strictly speaking also define the connected part of the correlator. For the scaling arguments presented here it is not necessary.

# 7  Discussion

In this paper we have presented arguments in favor of a spectral dimension equal two in quantum gravity as long as $c \leq 1$. In the same regime of central charge the Hausdorff dimension is measured to be close to four. These measurements are in agreement with recent independent numerical simulations for $0 < c < 1$, as well as somewhat older simulations for $c = -2$. In addition we have shown that both the Hausdorff dimension and the spectral dimension decrease for large $c$, while their ratio stays approximately equal to two.

The observation that the Hausdorff dimension, within the numerical accuracy of these simulations, is the double of the spectral dimension calls for a simple geometrical explanation. Does the fact that we have a non-zero probability for baby universe creation per unit area automatically imply $d_h = 2\bar{d}_s$?

The spectral dimension and the Hausdorff dimension seemingly test two different aspects of the dimensionality of an ensemble of manifolds. As argued in the introduction it is not so easy to understand why the spectral dimension of the ensemble of manifolds should be different from the dimension of the underlying manifold, while we do not have the same conceptional difficulties with the Hausdorff dimension, but the fact that both seem constant for $-\infty < c \leq 1$ indicates that both should be considered as *intrinsic properties of two-dimensional quantum gravity, independent of the coupling to matter*. The change in spectral dimension (and in Hausdorff dimension) for $c > 1$ thus indicates a drastic change in the theory, and it corroborates on the idea that for $c > 1$ (or at least for $c$ sufficiently large) the interaction between matter and gravity is so strong that the two-dimensional surface is torn apart. In the regularized, well-defined theory, described by dynamical triangulations, this seems to take place simply by a change in the ensemble of triangulated manifolds. The probability of picking a piecewise linear manifold, which is so degenerate (branched ?) that it does not qualify as a genuine two-dimensional manifold becomes one. This is seen in the computer simulations. The spectral dimension of all the individual triangulations decreases below two. However, it should be emphasized that the new theory for $c > 1$ seems not to be arbitrary. It keeps for instance the ratio between $d_h/\bar{d}_s \approx 2$. For large $c$ the theory probably degenerates to that of branched polymers, but there might be a region $1 < c < c_0$ where we have a non-trivial statistical theory. It is an interesting theoretical problem to understand this region.

---

[10]In [5] the spin-spin correlation function is measured, but no power law observed. The decay is fitted better to an exponential decay. It is presently difficult to give an interpretation to this result.

# 8  Appendix

## 8.1  Numerical calculation $G_V(R)$.

The Green function $G_V(R)$ is the inverse Laplace transform of $G_\Lambda(R)$, i.e.

$$G_V(R) = \int_{c-i\infty}^{c+i\infty} \frac{d\Lambda}{2\pi i} e^{\Lambda V} G_\Lambda(R) = \frac{1}{12\sqrt{\pi} V^{7/4}} U\left(\frac{R}{V^{1/4}}\right), \tag{79}$$

where $c$ is a positive real number and $U(x) = x^{d_h-1} F(x)$ is the dimensionless functions defined by (20) and (21). The constant $1/12\sqrt{\pi}$ is introduced to ensure the correct normalization (21) of $U(x)$:

$$U(x) = 12\sqrt{\pi} \int_{c-i\infty}^{c+i\infty} \frac{ds}{2\pi i} e^s s^{3/4} \frac{\cosh(xs^{1/4})}{\sinh^3(xs^{1/4})}. \tag{80}$$

The discretized version $n_N(r)$, as defined by (41), is related to $U(r/N^{1/4})$ in the limit of large $N$ by

$$n_N(r) = \frac{\alpha}{N^{1/4}} U\left(\frac{\alpha r}{N^{1/4}}\right), \tag{81}$$

where $\alpha$ is a constant parameter which depends on the regularization. In the case of the dynamical triangulations used here (which corresponds to one-matrix model with cubic potential) $\alpha = \sqrt{6/(12 + 13\sqrt{3})}$.

We now calculate the function $U(x)$ by performing numerically the inverse Laplace transformation. Four different methods are used:

**i)** direct numerical integration after $s$,

**ii)** analytic integration, after the mode expansion,

**iii)** saddle point integration, after the mode expansion,

**iv)** analytic integration after Taylor expansion around $x = 0$.

Method i) works well expect for $x$ close to 0 (i.e. in practice for $x > 1.8$), since a rescaling of $s$ with $1/x$ leads to the term $e^{s/x}$ in the integrand and it oscillates wildly for $x \to 0$.

Next, we consider the methods ii) and iii). The expansion

$$\cosh s / \sinh^3 s = 4 \sum_{n=1}^{\infty} n^2 \exp(-2ns)$$

can be given an interpretation as a "mode expansion" [1], and we can write

$$U(x) = 48\sqrt{\pi} \sum_{n=1}^{\infty} n^2 u(nx/2), \tag{82}$$

where

$$u(t) = \int_{c-i\infty}^{c+i\infty} \frac{ds}{2\pi i} \exp(s - 4ts^{1/4}) s^{3/4}. \tag{83}$$

Our task is now to evaluate (83) and next perform the summation (82).

It is possible to express $u(t)$ in terms of "known" functions since the inverse Laplace transform of $s^{a-1} \exp(-(bs)^{1/m})$ ($m = 1, 2, 3, \ldots$) can be expressed by the imaginary part of McRobert's $E$-functions. We have

$$u(t) = \frac{1}{\sqrt{2}\pi^{5/2}t^7} \Im[E(\frac{7}{4}, \frac{8}{4}, \frac{9}{4}, \frac{10}{4} : : e^{i\pi}t^4)]$$

$$= -\frac{3}{4\Gamma(1/4)} {}_1F_3(\frac{7}{4}; \frac{1}{4}, \frac{1}{2}, \frac{3}{4}; -t^4) + \frac{5\Gamma(1/4)t^2}{2\sqrt{2}\pi} {}_1F_3(\frac{9}{4}; \frac{3}{4}, \frac{5}{4}, \frac{3}{2}; -t^4)$$

$$- \frac{8t^3}{\sqrt{\pi}} {}_1F_3(\frac{5}{2}; \frac{5}{4}, \frac{3}{2}, \frac{7}{4}; -t^4), \tag{84}$$

where the hypergeometric function ${}_1F_3$ is defined by

$${}_1F_3(a; b_1, b_2, b_3; z) = \sum_{k=0}^{\infty} \frac{(a)_k}{(b_1)_k(b_2)_k(b_3)_k} \frac{z^k}{k!}, \tag{85}$$

and $(a)_k = \Gamma(a+k)/\Gamma(a)$.

In principle we can use (82),(84) and (85) to calculate $U(x)$ for all $x$. From a practical point of view it works well except for the smallest value of $x$, where we have to include very many terms in (82) and for very large $x$ where we have to include many terms in the hypergeometric series (85), i.e. in practice in a the region $0.02 < x < 10$.

For large values of $x$ it is sufficient to calculate the integral in (83), defining $u(t)$ by the saddle-point method iii). In this case we get:

$$u(t) = \sqrt{\frac{2}{3\pi}} e^{-3t^{4/3}} t^{5/3} \sum_{k=0}^{\infty} C_k \, t^{-4k/3}, \tag{86}$$

where

$$C_0 = 1, \quad C_1 = -\frac{115}{144}, \quad C_2 = -\frac{695}{41472}, \quad C_3 = -\frac{47755}{17915904},$$
$$C_4 = \frac{20518225}{10319560704}, \quad C_5 = -\frac{1316690375}{1486016741376}, \quad C_6 = -\frac{36815854075}{1283918464548864},$$
$$C_7 = \frac{162818713432375}{184884258895036416}, \quad C_8 = -\frac{366346837418687125}{21298666624708195 1232}, \quad \ldots \tag{87}$$

In general the saddle-point series is only an asymptotic series, but if we cut the series in (86) at $k = 5$ we get very good agreement with the use of the McRobert's $E$-functions for $x \in [1.6, 10]$. For large $x$ it is inconvenient to use the McRobert's $E$-functions, but for $x > 1.8$ we have good agreement agreement between the direct integration and the saddle-point approximation, which is of course the most convenient to use for $x \to \infty$.

The McRobert's $E$-functions becomes impractical for $x < 0.02$. For small $x$ we can turn to method iv). If we Taylor expand the integrand in (80) for small $x$ we get the following asymptotic expansion:

$$U(x) = -48x^3 \sum_{k=0}^{\infty} \frac{(\frac{5}{2})_k}{(\frac{5}{4})_k(\frac{3}{2})_k(\frac{7}{4})_k} \frac{\zeta(-4k-5)}{k!} (-\frac{x^4}{16})^k. \tag{88}$$

Eq. (88) is also obtained by substituting (85) into (82) with (84) and using the zeta-functional regularization, $\sum_{n=1}^{\infty} n^z = \zeta(-z)$. The convergence radius of the expansion is zero, but if we cut off the summation at $k$, and define this function as $U_k(x)$, we expect to get a good approximation for small $x$. By comparing with the results from the McRobert's $E$-function we get good agreement for $x \in [0.02, 1.6]$ if we use $U_9(x)$. Of course $U_9(x)$ gives an excellent approximation for $0 \le x < 0.02$ where we could not use easily the McRobert's $E$-functions. In this way we have managed to cover the whole interval $0 < x < \infty$, and except for $x \in [0, 0.02]$, by at least two independent approximations.

## 8.2 Asymptotic expansions in the general case

As discussed in the introduction the generalization of (79) and (80) is:

$$G_V(R) = \int_{c-i\infty}^{c+i\infty} \frac{d\Lambda}{2\pi i} e^{V\Lambda} G_\Lambda(R) \sim V^{\gamma_s - 1 - \nu} U(\frac{R}{V^\nu}), \qquad (89)$$

where $\nu = 1/d_h$ and

$$U(x) = \text{const.} \int_{c-i\infty}^{c+i\infty} \frac{ds}{2\pi i} e^s s^{\nu - \gamma_s} \hat{F}(xs^\nu). \qquad (90)$$

The leading term in the expansion of $\hat{F}(u)$ is $u^{\gamma_s/\nu - 1}$, i.e. the integrand starts with a term in $s$ which does not contribute to the Laplace transformation for finite volume (it gives a $\delta$-function in the volume). In fact we know that the first term which can contribute must be the term $u^{1/\nu - 1}$, which has to be present in the expansion. If we expand the integrand in (80) for small $x$, it contains odd powers of $x$. However, for half of these powers the function which multiplies $e^s$ is just an integer power of $s$, which does not contribute to the inverse Laplace transformation for finite area. This is the reason the expansion (88) jumps with powers of 4 in $x$. If we accept from the numerical experiments that the Hausdorff dimension $d_h = 4$ we have to use $\nu = 1/4$ in (90). If we at the same time use $\gamma_s = -1/m$, for the $(m, m+1)$ conformal field theory, the function $\hat{F}(u)$ has to be quite special. In fact it is natural to expect an expansion:

$$\hat{F}(u) = u^{\gamma_s/\nu - 1} \sum_{k=0}^{\infty} c_k (u^{-\gamma_s/\nu})^k \qquad (91)$$

where the first $k$ which gives a contribution to the inverse Laplace transformation for $V > 0$ is determined by the requirement that the power of $u$ should be $\nu - 1$ (in order to give $R^{d_h - 1}$), i.e.

$$\frac{\gamma_s}{\nu}(1-k) = \frac{1}{\nu}, \quad \text{i.e.} \quad k = m + 1. \qquad (92)$$

The above scenario is not very natural, but it is nevertheless realized in the case of pure gravity ($m = 2$). For $m \ne 2$ there are no obvious reasons to expect that the first correction term to $U(x)$ should be $x^7$ as is the case for $m = 2$ since the most naive choice of correction term from (91) appears to be $x^{3-\gamma_s/\nu}$.

There is another natural extension of pure gravity in the asymptotic form. We may expect

$$\hat{F}(u) = u^a \sum_{n=1}^{\infty} \rho(n) e^{-nu^b}. \qquad (93)$$

In the case of pure gravity, $a = 0$, $b = 1$, and $\rho(n) = n^2$. If $d_h = 4$ for any $m$, i.e., $U(x) \sim x^3 + \ldots$ for $x \approx 0$, we find that

$$a = 3(1-b), \quad b = \frac{2}{m}, \quad \text{and} \quad \rho(n) = \sum_{i=1}^{\infty} a_i n^{2i}, \qquad (94)$$

is a natural extension. Then, the leading term of $\hat{F}(u)$ becomes

$$\hat{F}(u) \sim u^{3(1-2/m)} \exp(-u^{2/m}). \tag{95}$$

After the inverse Laplace transformation we have

$$U(x) \sim \exp(-\text{const.} x^{4/(2m-1)}). \tag{96}$$

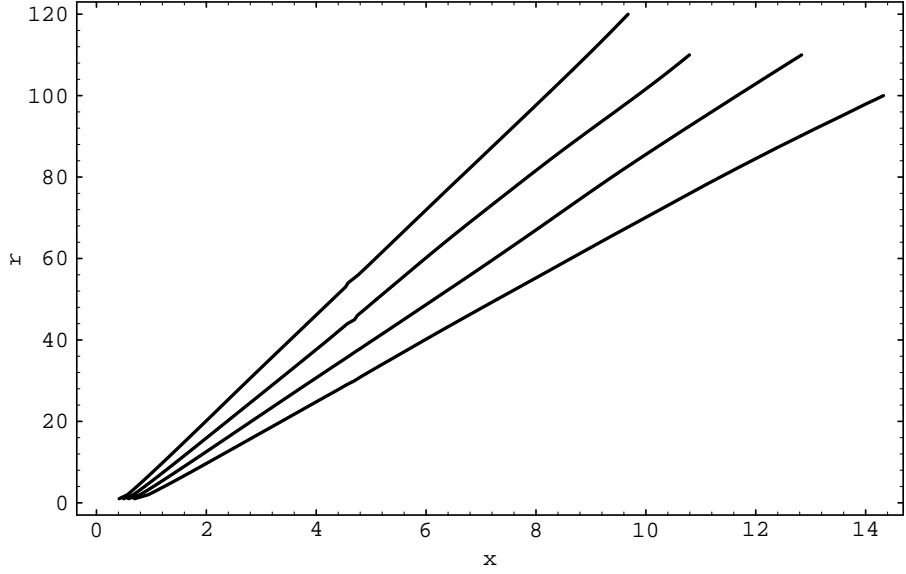

Figure 1: The "shift" in $r = cx - a$ as described by (43) for various size systems ($N =$ 4K, 8K, 16K and 32K) in pure gravity. The slope can be fitted to $c = N^{1/4} + b$, i.e. the 32K system corresponds to the graph with the steepest slope.

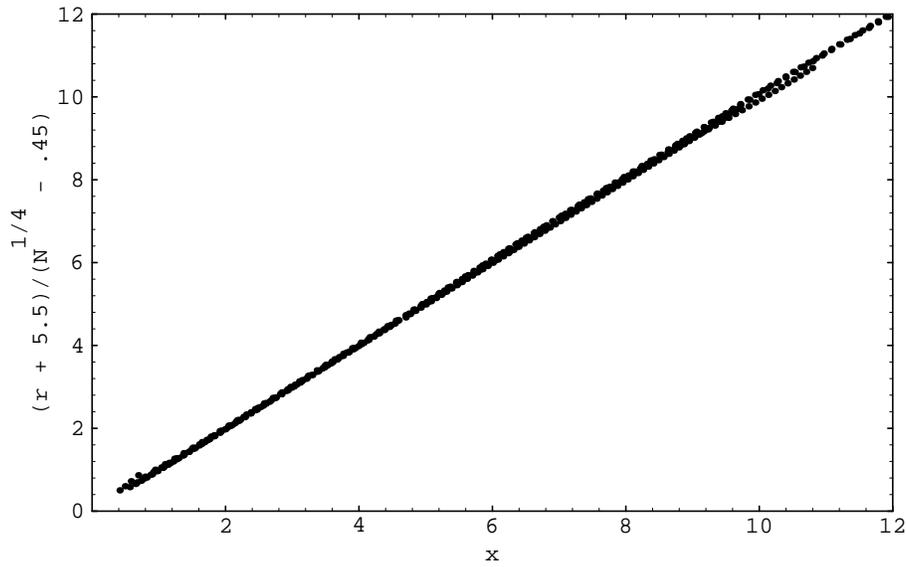

Figure 2: The function $(r + a)/(N^{1/4} + b)$, $a = 5.5$, $b = -.45$, plotted against $x$ (i.e. eq. (43)) for pure gravity and for $N = $ 1K, 2K, 4K,..., 32K.

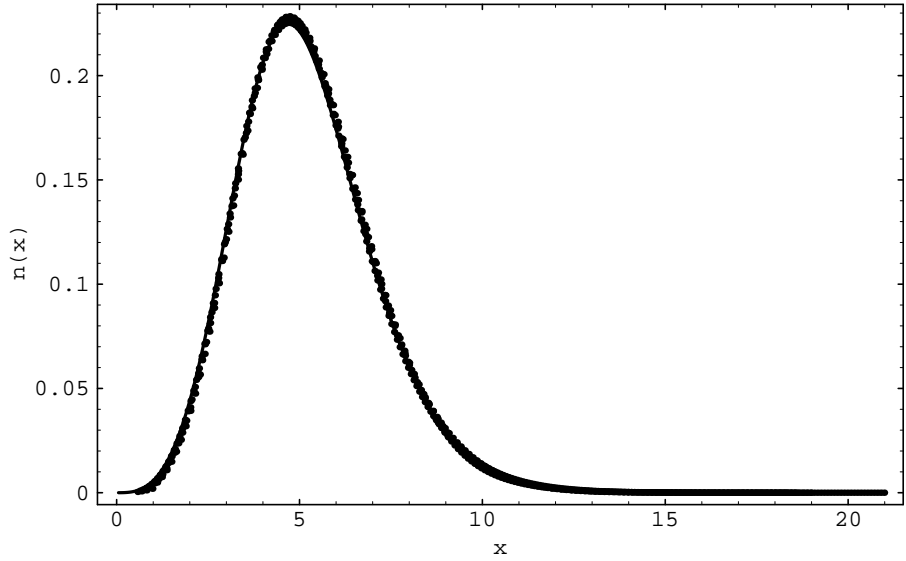

Figure 3: The scaled distributions for pure gravity for systems of sizes 1K, 2K, 4K,...., 32K triangles, as well as the theoretical distribution (the fully drawn line) as a function of the scaled variable $x = (r+5.5)/(N^{1/4} - 0.45)$.

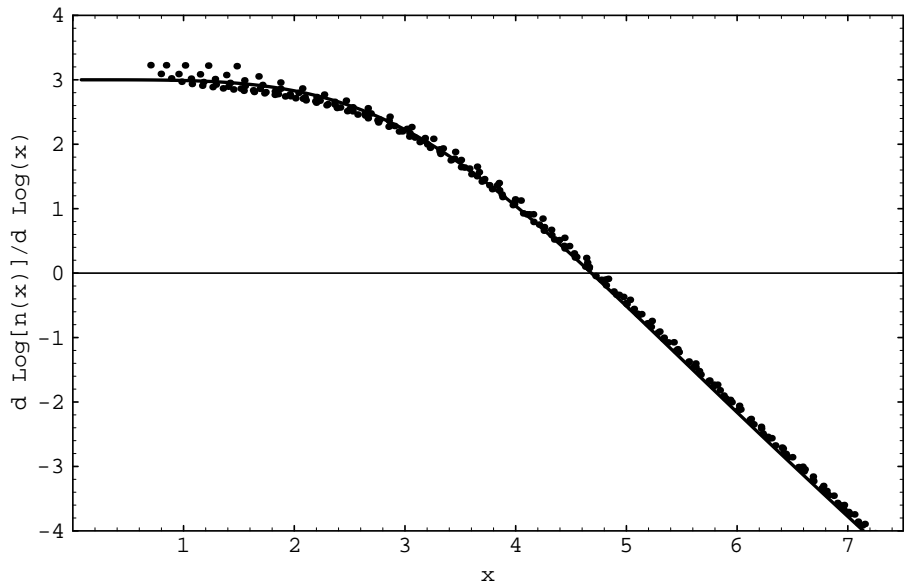

Figure 4: The derivative $dn(r(x))/dx$ plotted against $x$ for pure garivity. The same sizes of systems as in fig. 3 and again the fully drawn line is the theoretical curve.

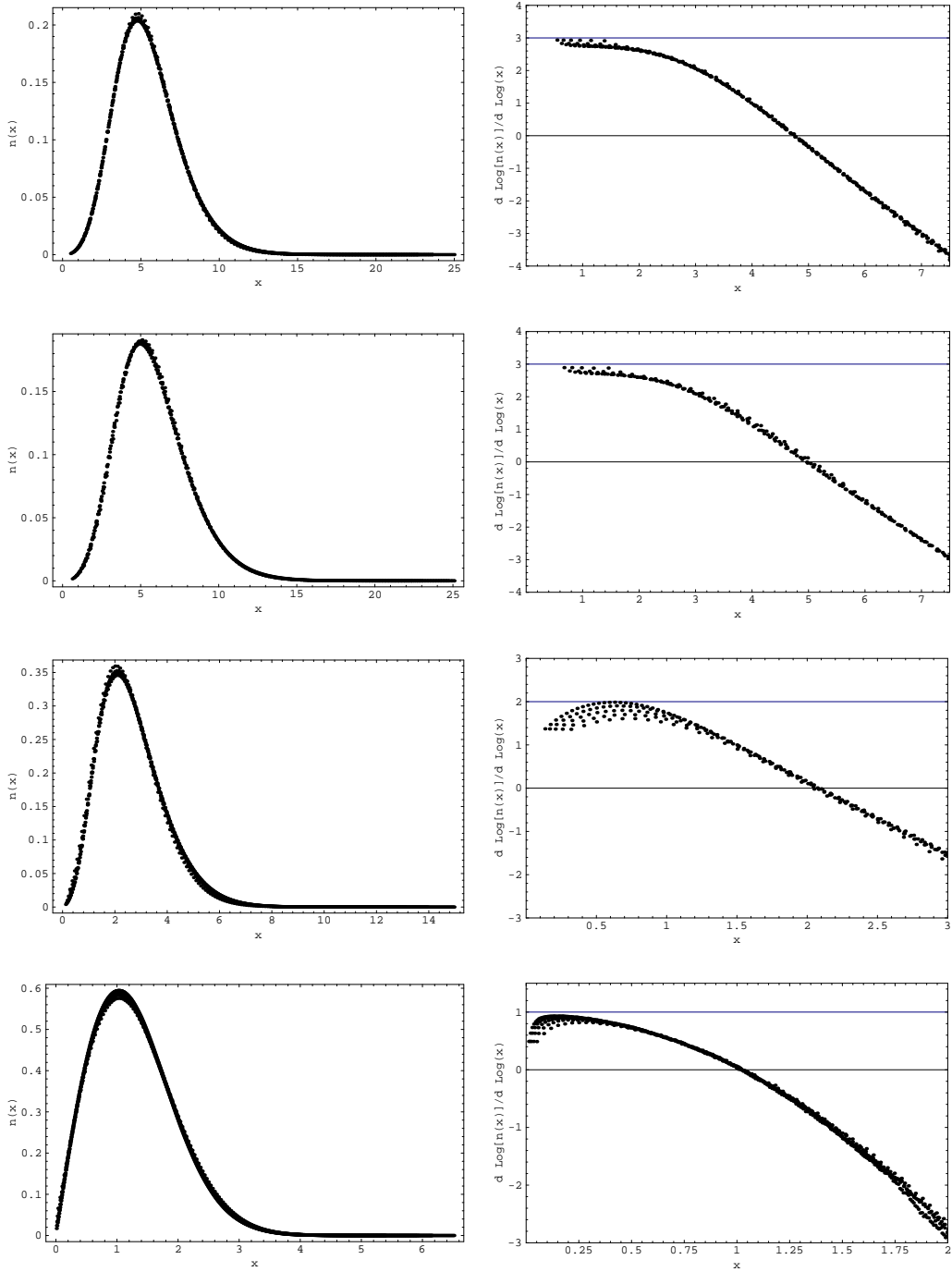

Figure 5: The scaled distributions $n(r(x))$ as well as the logarithmic derivative $d\log n(r(x))/dx$ for $c=$ 1/2, 1, 3 and 5. In constructing the mapping (43) we have assumed $d_h = 4$ for $c = 1/2$ and $c = 1$, $d_h = 3$ for $c = 3$ and $d_h = 2$ for $c = 5$. The constants $a, b$ determined this way are quite close to the pure gravity values for $c = 1/2$ and $c = 1$. In all cases the data include the following discretized volume sizes: $N = 1K$, 2K, 4K, ...,32K

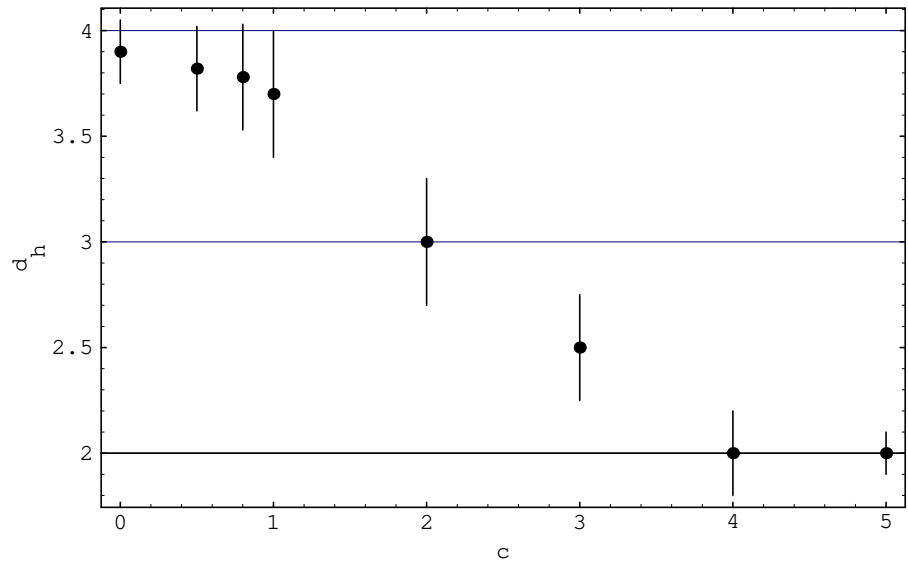

Figure 6: The Hausdorff dimension as determined by the finite size scaling relation for $c = 0$, $1/2$, $4/5$, $1$, $2$, $3$, $4$, and $5$. Contrary to fig. 5 $d_h$ has been treated as a free parameter in these fits and the dots denote the best values of $d_h$.

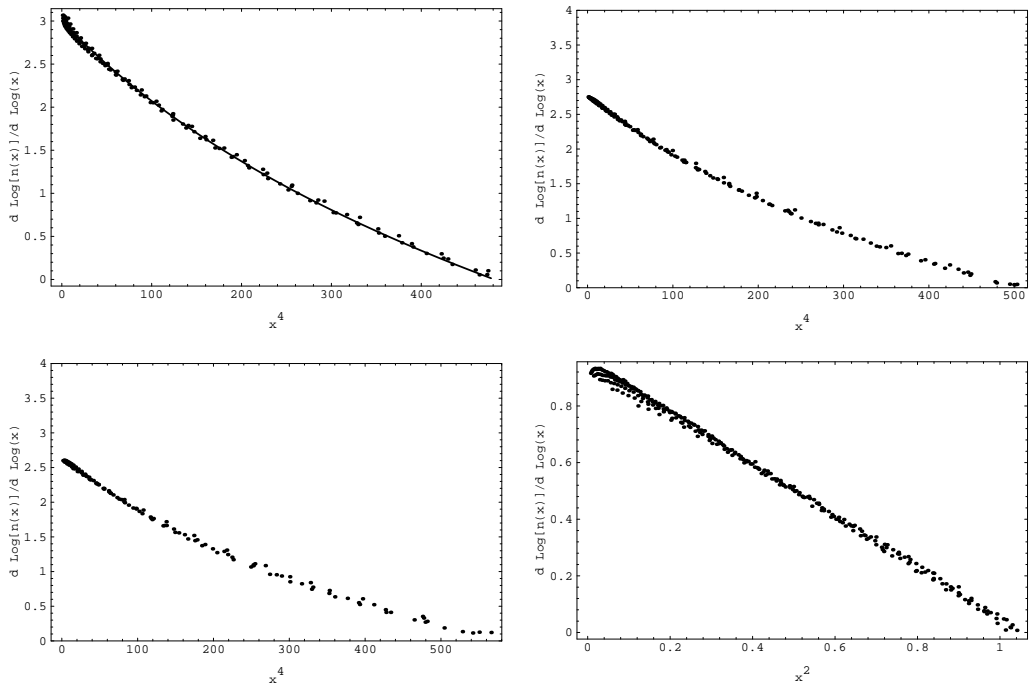

Figure 7: $d \log n(r(x))/dx$ as in (49) versus $x^\alpha$ for $c = 0$, $1/2$, 1 and 5. For $c \leq 1$, $\alpha \approx 4$ seems to give a well defined slope for $x \to 0$. For $c > 1$ it seems that $\alpha \approx 2$ if a slope at $x \to 0$ shall exist. All curves include volumes $N=$ 1K, 2K, 4K,...,32K.

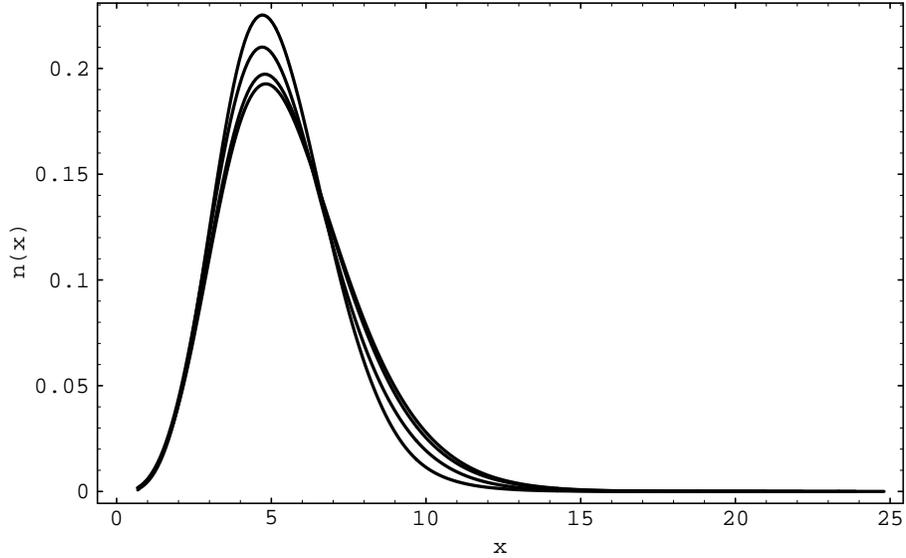

Figure 8: The scaled distribution for the various conformal theories with $c \leq 1$ coupled to gravity. The size of the systems is $N = 16000$. The values of $c$ is 0 (the highest peak) 1/2, 4/5 and 1 (the lowest peak). The parameters $d_h, a$ and $b$ used in (43) been chosen identical to the ones used in pure gravity ($d_h = 4$, $a = 5.5$ and $b = -0.45$), but even if these parameters are chosen from the best fit to the finite size scaling formulas the graphs shown will almost be unchanged. Note how well the results agree for small $x$ ($x \leq 3$).

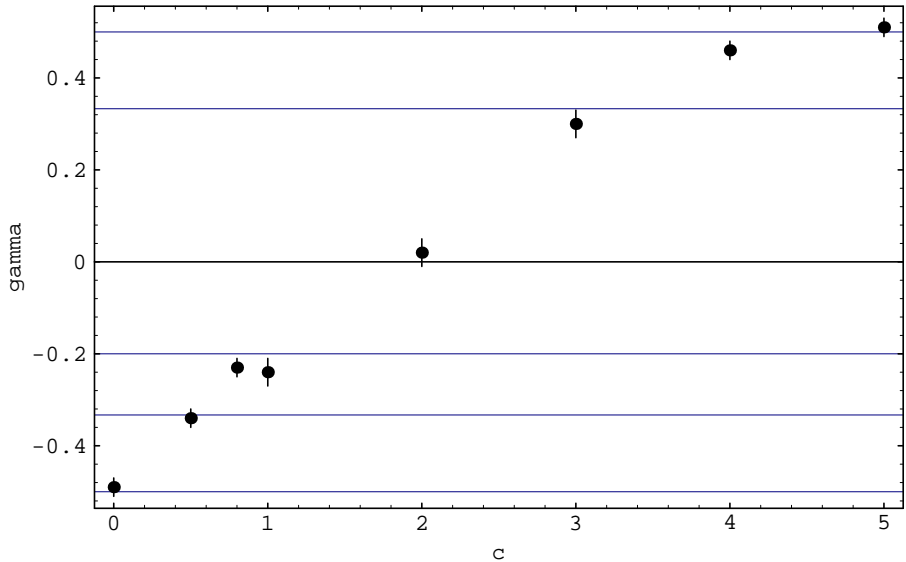

Figure 9: The measured string susceptibility versus central charge. No logarithmic corrections are used in the fits, even for $c = 1$. This is the reason the result for $c = 1$ is not the correct one (i.e. $\gamma_s(c = 1) = 0$).

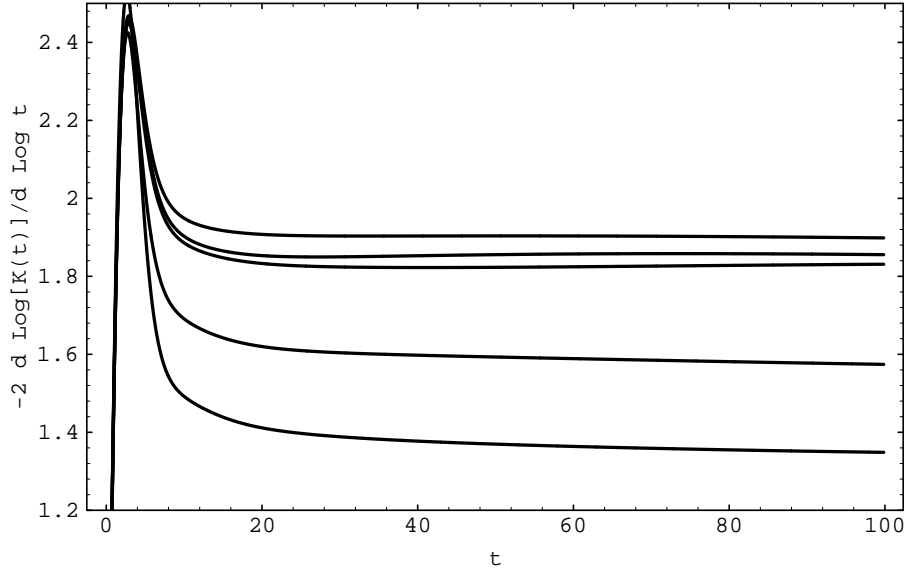

Figure 10: $-2d\log\mathrm{Tr}\,\hat{K}(t)/d\log t$ ($\approx \bar{d}_s$) versus $t$ for $c = 0$ (top curve), $c = 1/2$, $c = 1$, $c = 3$ and c=5 (bottom curve) theories coupled to 2d quantum gravity. The size of the systems is $N$=16K.

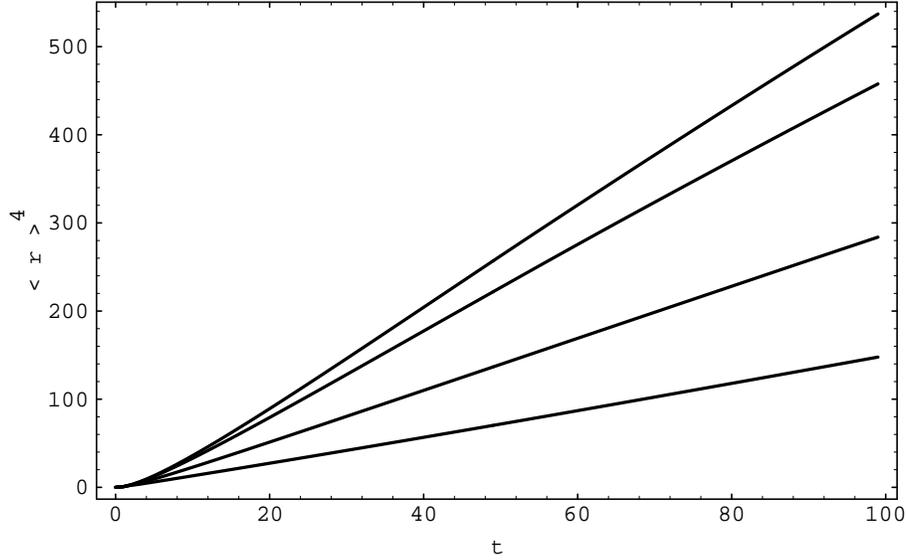

Figure 11: $\langle r(t) \rangle^4$ versus $t$ for various theories of matter coupled to gravity ($c$ =0 (bottom curve), $c = 1$, $c = 3$ and $c = 5$ (top curve)). The system size is $N = 4$K. Straight lines (as observed) indicate $2d_h/\bar{d}_s = 4$ according to (38)